\documentstyle[12pt]{article}
\newtheorem{theorem}{Theorem}[section]

\newtheorem{lemma}[theorem]{Lemma}

\newtheorem{remark}[theorem]{Remark}

\newcommand{\proof}{\noindent {\bf Proof. }}
\newcommand{\qed}{\hfill $\Box$ \vskip 2ex}
\def\range{{\rm Range}\ }

\newcounter{acount}

\newfont{\BB}{msbm10}
\def\R{\mbox{\BB R}}

\begin{document}

\date{\today}

\title{Generalized Hamilton-Jacobi equations for nonholonomic dynamics}
%%%%%%%%%%%%%%%%%%%%%%%%%%%%%%%%%%%%%%%%%%%%%%%%%%%%%%%%%%%%%%%%%%%%%%%%
\author{Michele Pavon\\Dipartimento di Matematica Pura e
Applicata\\Universit\`a of Padova\\via Belzoni 7, 35131 Padova\\and ISIB-CNR,
Italy\\pavon@math.unipd.it}
\date{\today}
%%%%%%%%%%%%%%%%%%%%%%%%%%%%%%%%%%%%%%%%%%%%%%%%%%%%%%%%%%%%%%%%%%%%%%%%
\maketitle
\begin{abstract} Employing a suitable nonlinear Lagrange functional, we derive generalized Hamilton-Jacobi equations for
dynamical systems subject to linear velocity constraints. As long as a
solution of the generalized Hamilton-Jacobi equation exists, the action
is actually minimized (not just extremized).
\end{abstract}

\noindent
{\bf PACS numbers:} 45.20.Jj, 45.10.Db

\noindent
{\bf Running Title:} Nonholonomic dynamics

\noindent
{\bf Keywords:} Nonholonomic systems, Hamilton-Jacobi equation, Hamilton's
principle

\section{Introduction} Consider a mechanical system with configuration space
$\R^n$. Let $L$ be the Lagrangian, and suppose that the system is subject
to
$k<n$ nonholonomic constraints of the form
\begin{equation}\label{PFFA}\omega_i(x(t))^T\dot{x}(t)=0,\quad
i=1,2,\ldots,k,\quad t\in [t_0,t_1],
\end{equation} where the $\omega_i:\R^n\rightarrow\R^n$ are smooth functions
and $T$ denotes transpose.  Let $\Omega(x)$ be the $k\times n$ matrix whose
$i^{th}$ row is $\omega_i(x)^T$. Then, an application of {\em
d'Alembert's principle}, together with the method of {\em Lagrange
undetermined multipliers}, gives that the equations of motion for the system
are
\begin{equation}\label{ELE}\frac{d}{dt}\frac{\partial L}{\partial
\dot{x}}-\frac{\partial L}{\partial x}=\Omega^T\lambda,
\end{equation}
where $\lambda$ is the $k$ dimensional Lagrange multiplier. Equation
(\ref{ELE}) together with (\ref{PFFA}) constitute a system of $n+k$ equations
for the $n+k$ unknowns
$x_1,x_2,\ldots,x_n,\lambda_1,\lambda_2,\ldots,\lambda_k$. The components of
$\Omega^T\lambda$ can be physically interpreted  as the components of the
(polygenic) force which acts on the mechanical system in order to maintain
the given non-holonomic conditions \cite{L}. Notice that d'Alembert's
principle is not variational. A variational approach to dynamics of
systems subject to linear velocity constraints was proposed in \cite{K}
(see also \cite[Chap. 1, Sect.4]{A}). A lucid critique of this ``Vakonomic
dynamics" (variational axiomatic kind dynamics) can be found in \cite{Za}. It is
shown there that the vakonomic equations may lead to paradoxical
behaviour. The relation between the vakonomic and holonomic approaches has also been discussed in \cite{A,KHA,LEW,CF,F,COR}.

We show in this paper that the second, hydrodynamic form of
Hamilton's principle may be extended to nonholonomic systems. As long as a
solution of the generalized Hamilton-Jacobi equation exists, the action
is minimized by a path satisfying the correct equations of motions
(\ref{ELE}). Our derivation relies on general nonlinear Lagrange functionals \cite{KP3,KP4}.
\section{The classical Hamilton principle}

We review below the hydrodynamic form of the classical Hamilton principle in
the strong form established in \cite[Section II]{P1} that developed from
\cite{GM}.
Consider a dynamical system with configuration space
$\R^n$. Let
\begin{equation}\label{LAG}L(x,v):=\frac{1}{2}mv\cdot v-V(x)
\end{equation}
be the Lagrangian function, where
$V(\cdot):\R^n\rightarrow
\R $ is of class $C^1$. Extension of the results of this paper to general Lagrangian functions that are strictly convex with respect to $v$ appears straightforward. We prefer, however, to treat the simple case (\ref{LAG}) in order to avoid obscuring ideas with technicalities. Let
${\cal X}_{0}$ denote the class of all $C^1$ paths
$x:[t_0,t_1]\rightarrow \R^n$ such that $x(t_0)=x_0$. Let
$\cal V$ denote the family of continuous functions
$v:[t_0,t_1]\rightarrow
\R^n$.  For $(x,v)\in ({\cal X}_{0}\times {\cal V})$, we define the
functional $J(x,v)$ by

\begin{equation}\label{H1}
J(x,v)=\int_{t_0}^{t_1}L(x(t),v(t))\,dt\:-\:S_1(x(t_1)),
\end{equation} where $S_1:\R^n\rightarrow\R$ is continuous. Consider the
following control problem:

\begin{equation}\label{PR} {\rm Minimize} \;\{J(x,v)| (x,v)\in ({\cal
X}_{0}\times {\cal V})\},
\end{equation} subject to the constraint
\begin{equation}\label{H2}
\dot{x}(t)=v(t),\quad \forall t\in [t_0,t_1].
\end{equation}
\begin{remark}It is apparent that this control problem is equivalent to
minimizing the action functional
$I(x):=J(x,\dot{x})$ over ${\cal X}_{0}$.
\end{remark}
To solve problem (\ref{PR})-(\ref{H2}) we
rely on the following elementary, albeit fundamental, result in the spirit of Lagrange.
Consider the  minimization of
$J:Y\rightarrow
\bar{\R}$, where
$\bar{\R}$ denotes the extended reals, over the nonempty subset
$M$ of $Y$.
\begin{lemma} {\em (Lagrange Lemma)}\label{L} Let $\Lambda :Y\rightarrow
\bar{\R}$ and let $y_0\in M$ minimize $J+\Lambda$ over $Y$. Assume that
$\Lambda(\cdot)$ is {\em finite} and {\em constant} over $M$. Then
$y_0$ minimizes $J$ over $M$.
\end{lemma}
\proof For any
$y\in M$, we have
$J(y_0)+\Lambda(y_0) \le J(y)+\Lambda(y) = J(y)+\Lambda(y_0)$. Hence
$J(y_0)\le J(y)$. \qed
\noindent A functional $\Lambda$ which is constant and finite on
$M$ is called {\it Lagrange functional}. For problem (\ref{PR}), let
$$M=\{(x,v)\in {\cal X}_0\times {\cal V}\, |\, \dot{x}(t)=v(t), \forall t\in
[t_0,t_1]\}.$$
We introduce a suitable  class of nonlinear Lagrange
functionals for our problem. Let
$F:[t_0,t_1]\times \R^n \rightarrow
\R$ be  of class $C^1$. Corresponding to such an $F$, we define the
nonlinear functional
$\Lambda^F$ on ${\cal X}_1\times {\cal V}$
$$
\Lambda^F(x,v):=F(t_1,x(t_1))-F(t_0,x(t_0))+\int_{t_0}^{t_1}
\left[-\frac{\partial F}{\partial t}(t,x(t))-v(t)\cdot \nabla
F(t,x(t))\right]dt.
$$ When
$(x,u)\in M$, by the chain rule, we have
$\Lambda^F(x,u)= 0$. Thus, $\Lambda^F$ is indeed a Lagrange functional for
our problem. The solution procedure is now outlined as follows.
\\{\bf Step 1.} Consider the {\it unconstrained}  minimization
\begin{equation} \label{PR2}                \min_{(x, v)\,\in \,({\cal
X}_1\times {\cal V})}  (J + \Lambda ^F)(x,v).
\end{equation}  We perform {\em two-stage} optimization. Namely, for each
fixed $x\in {\cal X}_0$ , we try to compute an optimal control
$v^*_x$ through {\it pointwise  minimization}  of the integrand of $J +
\Lambda^F$.  More explicitly, consider for each $x\in {\cal X}_0$  and each
$t\in [t_0,t_1]$ the finite-dimensional problem
\begin{equation}\label{R3}
\min_{v\in\R^n}\left\{\frac{1}{2}mv\cdot v- V(x(t)) -
\frac{\partial{F}}{\partial{t}}(t,x(t)) - v\cdot \nabla F(t,x(t))\right\}.
\end{equation}  We get
\begin{equation}\label{R4} v_x^*(t)=\frac{1}{m}\nabla F(t,x(t)).
\end{equation}  We notice that
$v^*_x$ belongs to the class of admissible velocities ${\cal V}$.
\\{\bf Step 2.}  Consider now the minimization of the functional
$$\Gamma^F(x)=(J + \Lambda^F)(x,v^*_x)$$  on the space ${\cal X}_0$. We
have
\begin{eqnarray}\nonumber
\Gamma^F(x)=&&-S_1(x(t_1))+F(t_1,x(t_1))-F(t_0,x(t_0))+\\
&&\int_{t_0}^{t_1}\left[-\frac{\partial{F}}{\partial{t}}(t,x(t))
-\frac{1}{2m}\nabla F(t,x(t))\cdot\nabla
F(t,x(t))-V(x(t))\right]\,dt.\nonumber
\end{eqnarray} If we can find
$S$ such that $\Gamma^{S}(\cdot)$ is actually  {\it constant} on $X_{0}$,
then {\em any pair} $(x,v^*_x)\in ({\cal X}_1\times{\cal V})$ solves problem
(\ref{PR2}). Then, by Lemma (\ref{L}),  if the pair $(x,v^*_x)$ satisfies
$$\dot{x}(t)=v^*_x(t)=\frac{1}{m}\nabla S(t,x(t)),\,\forall t\in[t_0,t_1],$$
it also solves the original constrained problem (\ref{PR})-(\ref{H2}).
\begin{theorem} {\rm (\cite{P1})} Let $S(t,x)$ be any $C^1$ solution on
$[t_0,t_1]\times\R^n$ of the terminal value problem
\begin{eqnarray}\label{H3} &&\frac{\partial{S}}{\partial{t}} +
\frac{1}{2m}\nabla {S} \cdot \nabla{S} + V(x) = 0,
\\ &&S(t_1,x)=S_1(x).\label{H4}
\end{eqnarray} Then any $x\in {\cal X}_{0}$ satisfying on $[t_0,t_1]$
\begin{equation}\label{H*}
\dot{x}(t)=\frac{1}{m}\nabla S(x(t),t),
\end{equation} solves together with $\frac{1}{m}\nabla S(x(t),t)$  problem
{\em (\ref{PR})-(\ref{H2})}.
\end{theorem}
\proof If $S$ solves (\ref{H3})-(\ref{H4}), we get
$\Gamma^{S}(x)\equiv S(x_0,t_0)$ on ${\cal X}_{0}$.
\qed
\noindent Notice that when a $C^1$ solution
$S(x,t)$ of (\ref{H3})-(\ref{H4}) exists, then there are also solutions $x$
of the differential equation (\ref{H*}) satisfying
$x(t_0)={x_0}$, and therefore optimal pairs. In this case, the action
functional is actually {\em minimized}, not just extremized. The difficulty
lies, of course, with the terminal value problem (\ref{H3})-(\ref{H4}) that,
in general, only has a local in
$t$ solution (namely, on some interval
$(\bar{t},t_1]$, $t_0<\bar{t}$).
\begin{remark} {\em Let us now assume that $S$ is of class $C^2$. 
Following \cite{GM}, let us introduce the
{\em acceleration field} $a(t,x)$ through a substantial time derivative
$$a(t,x):=\left[\frac{\partial }{\partial t}+\frac{1}{m}\nabla
S\cdot
\nabla\right] (\frac{1}{m}\nabla 
S)(t,x)=\frac{1}{m}\nabla\left[\frac{\partial{S}}{\partial{t}} +
\frac{1}{2m}\nabla {S} \cdot \nabla{S}\right](t,x).
$$
Then, (\ref{H3}) implies the local form of Newton's law
\begin{equation}\label{Newt}
a(t,x)=-\frac{1}{m}\nabla V(x).
\end{equation} }
\end{remark}

\section{Nonholonomic dynamical systems}

Consider a system subject to linear velocity constraints of the form
\begin{equation}\label{PF}
\Omega(x(t))\dot{x}(t)=0,\quad t\in [t_0,t_1],
\end{equation}  where $\Omega:\R^n\rightarrow \R^{k\times n}, k<n$ is a
continuous map. We assume that for eack $x\in \R^n$, the rows of $\Omega$ are
linearly independent. These constraints are called {\em Pfaffian}. A simple
example is provided by a disk rolling on a plane without slipping. More
complex nonholonomic systems with Pfaffian constraints occur in many problems
of robot motion planning and have therefore been the subject of intensive
study, see \cite{MLS,KBMM,BBCM} and references therein. Let
\begin{equation}\label{NHC2}
\Omega(x(t))v(t)=0,\quad t\in [t_0,t_1].
\end{equation}   We now study the control problem
(\ref{PR})-(\ref{H2})-(\ref{NHC2}), namely the same problem as in the
previous section when also constraint (\ref{NHC2}) is present. This problem
is equivalent to  minimizing the action functional
\begin{equation}\label{AF}I(x):=J(x,\dot{x})=\int_{t_0}^{t_1}L(x(t),\dot{x}(t))\,dt\:-\:S_1(x(t_1)),
\end{equation}
under the constraints (\ref{PF}). Reformulating the calculus of variations
problem as a control problem as before, we let
$$M=\{(x,v)\in {\cal X}_0\times {\cal V}\, |\,
\dot{x}(t)=v(t), \Omega(x(t))v(t)=0,\,\forall t\in [t_0,t_1]\}.$$ Let
$F:[t_0,t_1]\times \R^n \rightarrow \R$ be  of class $C^1$ and
$g:[t_0,t_1]\times \R^n \rightarrow \R^k$ be continuous. Corresponding to
such a pair, we define the nonlinear functional
$\Lambda^{F,g}$ on ${\cal X}_0\times {\cal V}$ by
\begin{eqnarray}\nonumber
&&\Lambda^{F,g}(x,v):=F(t_1,x(t_1))-F(t_0,x(t_0))+\\&&\int_{t_0}^{t_1}
\left[-\frac{\partial F}{\partial t}(t,x(t))-v(t)\cdot \nabla
F(t,x(t))+g(t,x(t))^T
\Omega(x(t))v(t)\right]dt.\nonumber
\end{eqnarray} It is apparent that $\Lambda^{F,g}$ is a Lagrange functional
for the problem since it is identically zero when (\ref{H2}) and (\ref{NHC2}) are satisfied. Following the same procedure as in the previous section, we consider the {\it
unconstrained}  minimization of $\Lambda^{F,g}(x,v)$ over $({\cal X}_0\times
{\cal V})$. For $x\in {\cal X}_0$ fixed, the pointwise minimization of the
integrand of $J +\Lambda^{F,g}$ at time $t$ gives
\begin{equation}\label{oc}  v_x^*(t)=\frac{1}{m}\left[\nabla
F(t,x(t))-\Omega^T(x(t))g(t,x(t))\right].
\end{equation} Notice that
$v^*_x\in {\cal V}$. We consider next the minimization of the functional
$$\Gamma^{F,g}(x)=(J+\Lambda^{F,g})(x,v^*_x)$$  on the space ${\cal X}_{0}$.
We have
\begin{eqnarray}\nonumber
\Gamma^{F,g}(x)=-S_1(x(t_1))+F(x(t_1),t_1)-F(x(t_0),t_0)\\
+\int_{t_0}^{t_1}\left[-\frac{\partial{F}}{\partial{t}}(t,x(t))
-\frac{1}{2m}||\nabla
F(t,x(t))-\Omega^T(x(t))g(t,x(t))||^2-V(x(t))\right]\,dt,\nonumber
\end{eqnarray}  where $||\cdot||$ denotes the Euclidean norm in $\R^n$. Let
$S(t,x)$ of class
$C^1$  and
$\mu(t,x)$ continuous solve on
$\R^n\times [t_0,t_1]$ of the initial value problem
\begin{eqnarray}\label{HJC} &&\frac{\partial{S}}{\partial{t}} +
\frac{1}{2m}||\nabla {S}-\Omega^T\mu||^2 + V(x) = 0,
\\ &&S(x,t_0)=S_0(x).\label{HJC2}
\end{eqnarray}  Then $\Gamma^{S,\mu}(x)\equiv S(x_0,t_0)$ on ${\cal X}_{0}$.
By Lemma (\ref{L}), if $x\in {\cal X}_{0}$ satisfies for all
$t\in [t_0,t_1]$
\begin{eqnarray}\label{S1}\dot{x}(t)=\frac{1}{m}\left[\nabla
S(t,x(t))-\Omega^T(x(t))\mu(t,x(t))\right],\\
\Omega(x(t))\dot{x}(t)=0,\label{S2}
\end{eqnarray} then it solves the problem together with the corresponding
feedback velocity (\ref{oc}).
\begin{remark} \label{NEW}{\em As in the unconstrained case, we now show
that (\ref{HJC}) implies the second principle of dynamics. Assume that
$S$  is of class $C^2$ and that
$\Omega$, and $\mu$ are of class $C^1$. The acceleration field is
again obtained through a
substantial derivative of the velocity field
\begin{eqnarray}\nonumber a(t,x):=\left[\frac{\partial }{\partial
t}+\frac{1}{m}\left[\nabla
S-\Omega^T\mu\right]\cdot
\nabla\right]
(\frac{1}{m}\left[\nabla
S-\Omega^T\mu\right])(t,x)\\=\frac{1}{m}\left\{\nabla\left[\frac{\partial{S}}{\partial{t}}
+\frac{1}{2m}||\nabla
{S}-\Omega^T\mu||^2\right]-\frac{\partial{(\Omega^T\mu)}}{\partial{t}}
\right\}(t,x).\nonumber
\end{eqnarray}}
{\em Then, (\ref{HJC}) yields}
\begin{equation}\label{Newt2}
a(t,x)=-\frac{1}{m}\nabla
V(x)-\frac{1}{m}\Omega^T\frac{\partial{\mu}}{\partial{t}}.
\end{equation}
{\em Define
$$\lambda(t,x):=-\frac{\partial{\mu}}{\partial t}(t,x).$$ For
$x\in {\cal X}_{0}$ satisfying (\ref{S1}) and of class $C^2$, let
$\lambda(t):=\lambda(t,x(t))$. We then get}
\begin{equation}\label{ELE2} m\, \ddot{x}(t)=-\nabla
V(x(t))+\Omega^T(x(t))\lambda(t),
\end{equation}
{\em namely equation (\ref{ELE}). By differentiating (\ref{S2}), a simple
calculation employing (\ref{ELE2}), shows that the Lagrange multipliers
$\lambda$ may be expressed as an instantaneous function of $x$ and
$\dot{x}$, see e.g. \cite[Section 2]{Za}, \cite[pp.269-270]{MLS}}.
\end{remark}
\vspace{.4 cm}
\noindent
Next we show how the multiplier $\mu$ can be eliminated in our
hydrodynamic context. If
$x$ satisfies (\ref{S1})-(\ref{S2}), then, plugging (\ref{S1}) into
(\ref{S2}), we get
$$\Omega(x(t))\frac{1}{m}\left[\nabla
S(t,x(t))-\Omega^T(x(t))\mu(t,x(t))\right]=0
$$ Since $\Omega$ has full row rank, the latter is equivalent to
\begin{equation}\nonumber\mu(t,x(t))=(\Omega(x(t))\Omega^T(x(t)))^{-1}
\Omega(x(t))
\nabla S(t,x(t))\label{mult}
\end{equation}  Plugging this into (\ref{S1}), we get
\begin{equation}\label{S3}\dot{x}(t)=\frac{1}{m}\left[\left(I-
\pi(x(t))\right)\nabla S(t,x(t))\right],
\end{equation}
where $\pi (t,x)$ is defined by
\begin{equation}\label{PRO}\pi(x)=\Omega^T(x)(\Omega(x)\Omega^T(x))^{-
1}\Omega(x).
\end{equation} Observe that $\pi(x)^2=\pi(x)$ and $\pi(x)^T=\pi(x)$. Thus,
$\pi(x)$ is an orthogonal projection. In fact, $\pi(x)$ is the orthogonal
projection onto $\range (\Omega^T(x))$.

\begin{remark}\label{re}
{\em Notice that (\ref{S3}) implies
(\ref{S2}). Indeed,
$$\Omega(x(t))\dot{x}(t)=\Omega(x(t))\frac{1}{m}\left[\left(I-
\pi(x(t))\right)\nabla S(t,x(t))\right]=0,
$$
since $I-\pi(x)$ projects onto the kernel of $\Omega(x)$.
}
\end{remark}
Now we use the freedom we have in picking $S$ and
$\mu$. Since (\ref{mult}) must be satisfied by an optimal solution, we impose
that the pair $(S,\mu)$ satisfies identically on all of
$[t_0,t_1]\times\R^n$
\begin{equation}\label{mu}\mu(t,x)=(\Omega(x)\Omega^T(x))^{-1}\Omega(x)
\nabla S(t,x).
\end{equation} Hence, $\Gamma^{S,\mu}(x)=\Gamma^{S}(x)$ and equation
(\ref{HJC}) becomes
\begin{equation}\label{GHJB}\frac{\partial{S}}{\partial{t}} +
\frac{1}{2m}||(I-\pi)\nabla {S}||^2 + V(x) = 0
\end{equation}
We are now ready for our main result.
\begin{theorem}\label{THEO} For $x\in \R^n$, let $\sigma(x)=I-\pi(x)$ denote
the orthogonal projection onto $\ker \Omega(x)$. Let $S(t,x)$ be any
$C^1$ solution on
$[t_0,t_1]\times\R^n$ of
\begin{equation}\label{GHJB2}\frac{\partial{S}}{\partial{t}} +
\frac{1}{2m}\nabla S\cdot\sigma\nabla {S}+ V(x) = 0,\quad
S(t_1,x)=S_1(x).
\end{equation} Then any $x\in {\cal X}_{0}$ satisfying
\begin{equation}\label{EQ}\dot{x}(t)=\frac{1}{m}\sigma(x(t))\nabla S(t,x(t))
\end{equation} on $[t_0,t_1]$ solves together with
$$v(t)=\frac{1}{m}\sigma(x(t))\nabla S(t,x(t))
$$  problem (\ref{PR})-(\ref{H2})-(\ref{NHC2}) (equivalently, such an $x\in  {\cal X}_{0}$ minimizes (\ref{AF}) subject to (\ref{PF})). If $S$ is of class
$C^2$, and $x\in {\cal X}_{0}$ satisfying (\ref{EQ}) is also of class
$C^2$, then $x$ satisfies equation (\ref{ELE2})
$$ m\, \ddot{x}(t)=-\nabla
V(x(t))+\Omega^T(x(t))\lambda(t),
$$
with
$\lambda$ given by
\begin{equation}\label{mul}
\lambda(t)=-(\Omega(x(t))\Omega^T(x(t)))^{-1}\Omega(x(t))\nabla
\frac{\partial S}{\partial{t}}(t,x(t)).
\end{equation}
\end{theorem}
\proof If $S$ solves (\ref{GHJB2}), we get
$\Gamma^{S,\mu}(x)=\Gamma^{S}(x)=S(x_0,t_0)$ for any $x\in {\cal X}_{0}$.
Thus any pair
$(x,v)\in ({\cal X}_0\times {\cal V})$ solves the unconstrained minimization
problem. Since $x$ satisfies (\ref{EQ}), constraint (\ref{H2}) is fulfilled.
Moreover, by Remark \ref{re},
(\ref{NHC2}) is also satisfied. By Lemma \ref{L}, the pair is optimal for the
original constrained minimization. Finally, $x$ satisfies (\ref{ELE2})
with $\lambda$ as in (\ref{mul}) in view of Remark \ref{NEW} and
(\ref{mu}).
\qed
\noindent{\bf Acknowledgments} I wish to thank Luis Bonilla and Franco
Cardin for some  useful conversations on the content of this paper and for
pointing out some relevant bibliography to me.

\end{document}